\documentclass[seceq]{ptptex}

\usepackage{graphicx}

\title{
The in-medium $\bar K N$ interaction within a chiral unitary approach%
}

\author{
Laura \textsc{Tol\'os}$^{1,}$\footnote{ e-mail address:
tolos@th.physik.uni-frankfurt.de},  
Angels \textsc{Ramos}$^{2,}$\footnote{ e-mail address:
ramos@ecm.ub.es} 
and
Eulogio \textsc{Oset}$^{3,}$\footnote{ e-mail address:
oset@ific.uv.es} 
}

\inst{
$^1$ FIAS, J.W. Goethe-Universit\"at, Max-von-Laue 1, 60438 Frankfurt (M), Germany\\
$^2$Departament d'Estructura i Constituents de la Mat\`{e}ria,
     Universitat de Barcelona,
     Diagonal 647, 08028 Barcelona, Spain  \\
$^3$ Departamento de F\'{\i}sica Te\'orica and IFIC, 
     Centro Mixto Universidad de Valencia-CSIC,
     Institutos de Investigaci\'on de Paterna, 
     Ap. Correos 22085, E-46071 Valencia, Spain 
}

\abst{ The $s$- and $p$-wave contributions to the  $\bar K N$ interaction in dense nuclear matter are obtained using a chiral unitary approach. We perform a self-consistent calculation of the $\bar K$ self-energy including Pauli blocking effects, meson self-energies modified by short-range correlations and baryon binding potentials. We find that the on-shell factorization cannot be applied to evaluate the in-medium corrections to $p$-wave amplitudes. Furthermore, the $\Lambda$ and $\Sigma$ develop a mass shift of -30 MeV at saturation density  while
the $\Sigma^*$ width increases to  80 MeV. We conclude that no deep and narrow $\bar K$ bound states could be observed.

} 
\begin{document}

\maketitle
Phenomenology of kaonic atoms shows that the $\bar K$ feels an attractive potential at low densities. This attraction results from the modified $s$-wave $\Lambda(1405)$ resonance in the medium  \cite{review} due to Pauli blocking effects \cite{koch} combined with the self-consistent consideration of the $\bar K$ self-energy \cite{lutz,schaffner} and the inclusion of self-energies of the mesons and baryons in the intermediate states \cite{Ramos:1999ku}. Attraction of the order of -50 MeV at normal nuclear matter density $\rho_0=0.17 \,{\rm fm^{-3}}$ is obtained by different approaches, such as unitarizated extensions of chiral theories in coupled-channels \cite{Ramos:1999ku}. 

Further studies of higher-partial waves have been performed recently \cite{laura,lutz-korpa02}. The $p$-wave contribution to the $\bar K N$ optical potential has been found to be negligible for atoms  \cite{angelscarmen}. However, heavy-ion collisions can test high-momentum kaons and, therefore, further partial-wave contributions.

In this paper we study the $\bar K$ properties in dense  matter using a chiral unitary approach which incorporates  $s$- and $p$-wave contributions to the $\bar K N$ interaction. We show that the on-shell factorization cannot be applied for $p$-waves in the medium. The self-energy of $\Lambda(1115)$, $\Lambda(1405)$, $\Sigma(1195)$ and $\Sigma^*(1385)$ is also analyzed \cite{tolos}.

\begin{figure}[hbt]
\begin{minipage}{10cm}
  \includegraphics[height=5cm,width=0.65\textwidth]{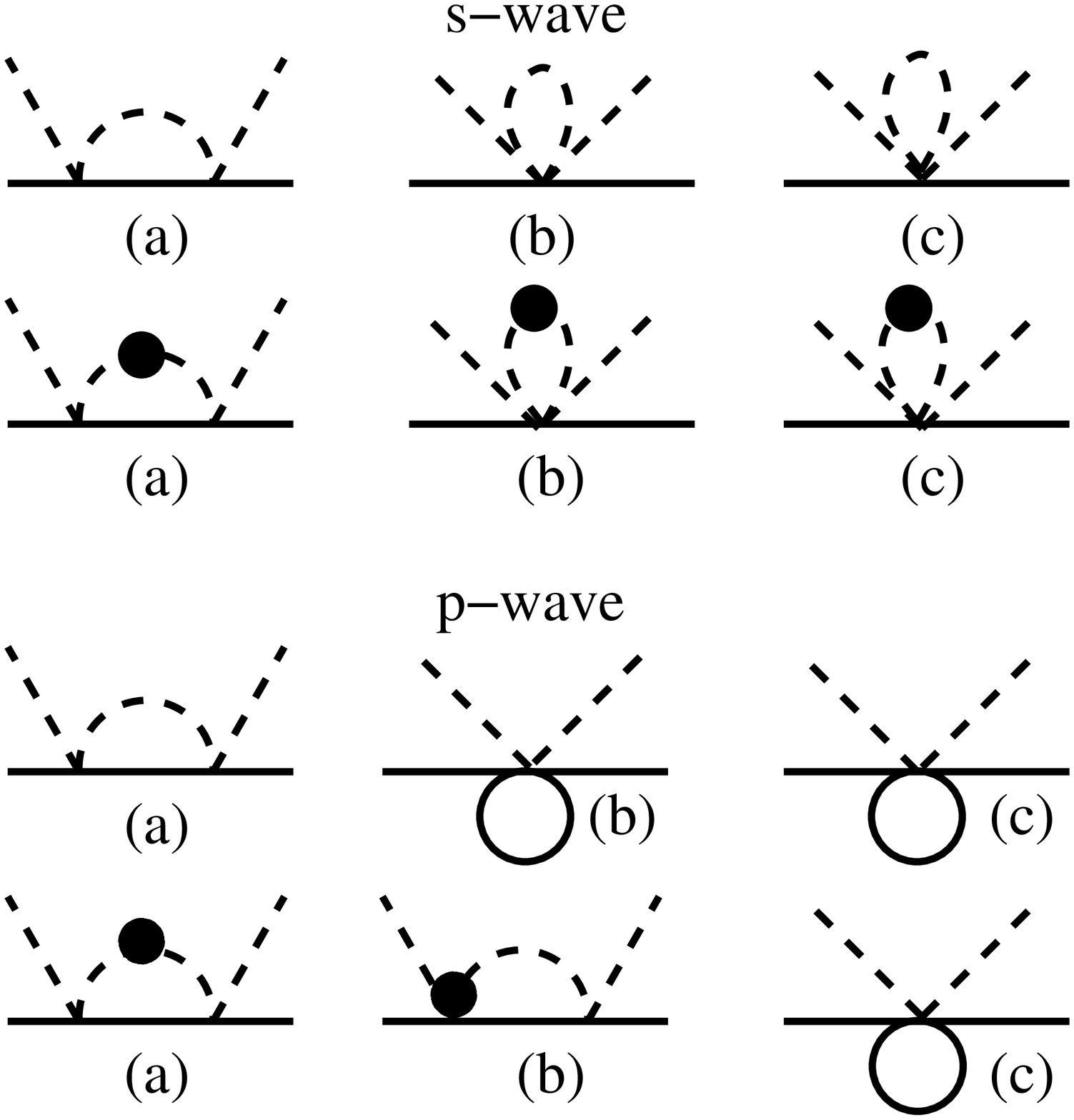}
  \includegraphics[width=0.7\textwidth]{lambda-sigma.eps}
\end{minipage}
\caption{Left: On-shell (a), off-shell (b) and tadpole (c) contributions for $s$- (upper two rows) and $p$-wave (lower two rows) in free space and with self-energy insertions. Right: $\Lambda(1115)$, $\Lambda(1405)$, $\Sigma(1195)$ and $\Sigma^*(1385)$ resonances.}
\label{fig:1}       
\end{figure}

The $\bar K$ self-energy in the nuclear medium and, hence, the spectral density  are obtained by incorporating medium modifications to the coupled-channel Bethe-Salpeter equation using, as  kernel, tree level chiral contributions. The $s$-wave contribution is derived from the lowest-order chiral lagrangian that couples the octet of pseudoscalar mesons to the octet of $1/2^+$ baryons \cite{oset}, while the $p$-wave amplitudes come mainly from the $\Lambda$, $\Sigma$ and $\Sigma^*$ pole terms \cite{jido}. For meson-baryon scattering the kernel can be factorized on the mass shell in the loop functions \cite{oset,Oller:2000fj}. The loop function is then regularized by means of a cutoff or dimensional regularization. The formal result is $T=[1-VG]^{-1}V$, where $V$ is the kernel and $G$ the loop function. 

The medium modifications arise from the inclusion of Pauli blocking effects on the nucleons
and the dressing of mesons and baryons in the intermediate loops. The binding effects for baryons are taken within the mean-field approach \cite{tolos}. For $\bar K$ and pions the medium modifications are included via the corresponding self-energy \cite{Ramos:1999ku}.

The in-medium amplitudes are obtained using a similar unitarization procedure as in free space. However, the on-shell factorization of the kernel out of the loop function is  valid for in-medium $s$-wave amplitudes while not for $p$-wave contributions.

In the l.h.s. of Fig.~\ref{fig:1}, the off-shell contributions for $s$-wave of the two vertices in the loop function eliminates  a baryon propagator. Hence, this off-shell term ((b) in row 1) is cancelled by the presence of a tadpole term ((c) in row 1), in a suitable renormalization scheme. By incorporating self-energy insertions in the meson line, the cancellation between the off-shell part ((b) in row 2) and the tadpole ((c) in row 2) still holds. The in-medium $s$-wave amplitudes are thus obtained by solving the Bethe-Salpeter equation with on-shell amplitudes and the in-medium loop functions.

\begin{figure}[hbt]
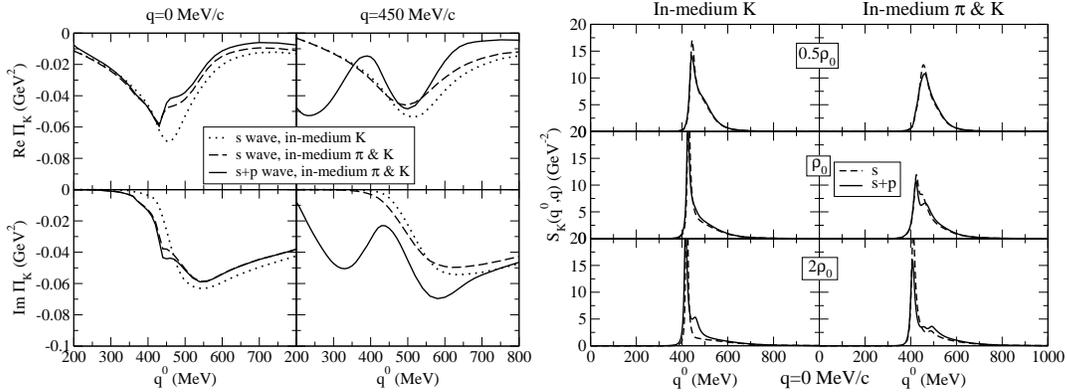

\begin{minipage}{10cm}
  \includegraphics[width=0.7\textwidth]{self.eps}
  \includegraphics[width=0.7\textwidth]{spec.eps}
\end{minipage}
\caption{Left: $\bar K$ self-energy as function of $\bar K$ energy for two different kaon momenta. Right: $\bar K$ spectral function  as function of $\bar K$ energy for zero momentum for two approaches. }
\label{fig:2}       
\end{figure}

The situation for the $p$-wave amplitudes is, however, different (see l.h.s of Fig.~\ref{fig:1}). The off-shell part ((b) in row 3) eliminates a meson propagator \cite{Cabrera:2002hc} and, then, it can be cancelled by the tadpole term ((c) in row 3).  However, when the meson propagator is dressed, the off-shell term ((b) in row 4) cancels only one of the two intermediate meson propagators. Hence, in the medium, we do not find the cancellation between the off-shell part  and tadpole term ((c) in row 4). This is solved adding to the free loop the medium corrections using the full off-shell $\vec{q}\,^2$ dependence 
\vspace{-0.2cm}
\begin{eqnarray}
\label{gtogmedio}
&&G_l^p(s) \to G_l^p(s) + \frac{1}{\vec{q}\,^2_{on}} \lbrack I_{\rm med}(s) -
I_{\rm free}(s)
\rbrack \ ,\nonumber\\
&&I_{\rm med}(s) = i \int \frac{d^{4}q}{(2\pi)^{4}} \vec{q}\,^2 D_M(q) G_B(P-q) \nonumber \\
&& I_{\rm free}(s) = i \int \frac{d^{4}q}{(2\pi)^{4}} \vec{q}\,^2 D^0_{M}(q)
G^0_{B}(P-q) \ ,
\end{eqnarray}
with $D_M$ and $G_B$ being the meson and baryon propagators and $P=q+p$, where $q$ and $p$ are the meson and baryon four-momentum in the lab frame. Another ingredient to be considered when dealing with in-medium $p$-wave amplitudes is the inclusion of short-range correlations (see details in Ref.~\cite{tolos}). The in-medium $p$-wave amplitudes are then obtained from the Bethe-Salpeter equation using the in-medium meson-baryon propagators of Eq.~(\ref{gtogmedio}), which incorporate the right $\vec{q}\,^2$ dependence, as well as Pauli blocking effects, dressing of mesons and baryons,  and short-range correlations.

The $\bar K$ self-energy is calculated self-consistently summing the in-medium $\bar K N$ interaction  $T_{\bar K N}$ for $s$- and $p$-waves over the Fermi sea of nucleons $n(\vec{p})$ 
\begin{eqnarray}
\Pi_{\bar{K}}(q^0,{\vec q},\rho)=\int \frac{d^3p}{(2\pi)^3}\,
n(\vec{p}\,) \, (T_{\bar K N}(I=0)\, + \, 3\, T_{\bar K N}(I=1))(P^0,\vec{P},\rho) \ , 
\end{eqnarray}
and the $\bar K$ spectral function is then given by
\begin{eqnarray}
S_{\bar K}(q^0,{\vec
q},\rho)= -\frac{1}{\pi}\frac{{\rm Im} \Pi_{\bar K}(q^0,\vec{q},\rho)}
{\mid (q^0)^2-\vec{q}\,^2-m_{\bar K}^2-
\Pi_{\bar K} (q^0,\vec{q},\rho) \mid^2} \ .
\end{eqnarray}

In the r.h.s. of Fig.~\ref{fig:1} we display the results for the $\Lambda(1115)$, $\Lambda(1405)$, $\Sigma(1195)$ and $\Sigma^*(1385)$ resonances. The free amplitudes (dotted lines) are compared to the in-medium ones at  $\rho_0$ dressing the antikaons self-consistently (dashed lines) and also considering  the in-medium effects on pions (solid lines). The $\Lambda(1115)$ acquires an attractive shift of -28 MeV when pions are dressed due to the appearance of the pion-mediated $\Lambda N \to \Sigma N$ channels. This is in agreement with hypernuclear spectroscopy data \cite{hyper}. The $\Lambda(1405)$ is generated dynamically close to the free position and gets strongly diluted when the in-medium properties of pions are considered due to $\Lambda N N^{-1}$ and $\Sigma N N^{-1}$ excitations. The  $\Sigma(1195)$ shows an attraction of -35 MeV when pions are dressed in line with \cite{Batty:1978sb} and in contrast with -10 MeV \cite{lutz-korpa02} or even the repulsion in Ref~\cite{Kaiser:2005tu}. 
The only reliable evidence, obtained from kaonic atoms, is that the $\Sigma(1195)$ requires attraction at small densities. The $\Sigma^*(1385)$ stays close to the free position for both approaches, in contrast to Ref.~\cite{lutz-korpa02}, and increases the width from 30 MeV in free space to 80 MeV when pions are dressed.

The self-energy of $\bar K$ at $\rho_0$ as function of energy is shown in the l.h.s. of Fig.~\ref{fig:2} for two momenta. We show the $s$-wave component only dressing kaons (dotted lines), dressing also pions (dashed lines) and the $s$- and $p$-wave contributions for the latest approach (solid lines).  The small $p$-wave strength at $q=0 \, {\rm MeV/c}$ is due the Fermi motion of nucleons which  produces a slight repulsion in the real part of the self-energy since the energies that come into play are above the
 $\Lambda$, $\Sigma$ and $\Sigma^*$ excitations. At finite momentum of 450 MeV/c, the imaginary part shows the $\Sigma N^{-1}$ component at 300 MeV and the $\Sigma^* N^{-1}$ one around 550 MeV. The corresponding optical potential at the quasiparticle energy changes from  -30 MeV to -80 MeV in the range of $\rho_0/2$ to $2 \rho_0$ when pions are dressed. A similar shift is obtained when only $\bar K$ are dressed self-consistently while the imaginary part is sizeable for both approaches.

The antikaon spectral function at $q=0 \, {\rm MeV/c}$ is displayed in the r.h.s. of Fig.~\ref{fig:2}. The spectral function does not show a Breit-Wigner behaviour. The slow fall off on the right-hand side of the quasiparticle peak is due to the $\Lambda(1405)N^{-1}$ excitation and the $p$-wave components are the result of the Fermi motion of nucleons. With increasing density, the quasiparticle peak gains attraction and the spectral function is diluted. The small peak observed on the right-hand side of the quasiparticle peak at $2\rho_0$ is due to the $\Sigma^*(1385)N^{-1}$ excitation.

In summary, we have obtained the $\bar K$ self-energy within a chiral unitary approach and, as a byproduct, the properties of the $\Lambda$, $\Sigma$ and $\Sigma^*$ hyperons in nuclear matter. While the $\Lambda$ and  $\Sigma$ feel an attractive potential of -30 MeV at $\rho_0$, the $\Sigma^*$ barely changes its mass but develops a width of 80 MeV. According to the $\bar K$ self-energy obtained, we conclude that no deep and bound $\bar K$ states could be observed. 

\section*{Acknowledgements}
The authors thank the Yukawa Institute for Theoretical Physics at Kyoto University. Discussions during the YKIS2006 on "New Frontiers on QCD'' were useful to this work.
L.T. acknowledges support from Gesellschaft f\"ur Schwerionenforschung. 



\begin{thebibliography}{}
%
\bibitem{review}
  J.~A.~Oller, E.~Oset and A.~Ramos,
  Prog.\ Part.\ Nucl.\ Phys.\  {\bf 45} (2000), 157.
\bibitem{koch}
  V.~Koch,
  Phys.\ Lett.\ B {\bf 337} (1994), 7. 
\bibitem{lutz}
  M.~Lutz,
  Phys.\ Lett.\ B {\bf 426}  (1998), 12.
\bibitem{schaffner}
  J.~Schaffner-Bielich, V.~Koch and M.~Effenberger,
  Nucl.\ Phys.\ A {\bf 669} (2000), 153.
\bibitem{Ramos:1999ku}
A.~Ramos and E.~Oset,
Nucl.\ Phys.\ A {\bf 671} (2000), 481 .
\bibitem{laura}
  L.~Tolos, A.~Ramos, A.~Polls and T.~T.~S.~Kuo,
  Nucl.\ Phys.\ A {\bf 690} (2001), 547.
\bibitem{lutz-korpa02}
  M.~F.~M.~Lutz and C.~L.~Korpa,
  Nucl.\ Phys.\ A {\bf 700} (2002), 309.
\bibitem{angelscarmen}
  C.~Garcia-Recio, E.~Oset, A.~Ramos and J.~Nieves,
  Nucl.\ Phys.\ A {\bf 703} (2002), 271. 
\bibitem{tolos}
  L.~Tolos, A.~Ramos and E.~Oset,
  Phys.\ Rev.\ C {\bf 74} (2006), 015203
\bibitem{oset}
E.~Oset and A.~Ramos,
Nucl.\ Phys.\ A {\bf 635} (1998), 99.
\bibitem{jido}
D.~Jido, E.~Oset and A.~Ramos,
Phys.\ Rev.\ C {\bf 66} (2002), 055203.
\bibitem{Oller:2000fj}
J.~A.~Oller and U.~G.~Meissner,
Phys.\ Lett.\ B {\bf 500} (2001), 263.
\bibitem{Cabrera:2002hc}
D.~Cabrera and M.~J.~Vicente Vacas,
Phys.\ Rev.\ C {\bf 67} (2003), 045203.
\bibitem{hyper} T. Hasegawa et al., Phys. Rev. C {\bf 53} (1996),
1210; P.H. Pile et al., Phys. Rev. Lett. {\bf 66} (1991),
2585; D.H. Davis, J.Pniewski, Contemp. Phys. {\bf 27} (1986), 91;
M. May et al., Phys. Rev. Lett. {\bf 78} (1997), 4343.
\bibitem{Batty:1978sb}
C. J. Batty et al.,
Phys.\ Lett.\ B {\bf 74} (1978), 27.
\bibitem{Kaiser:2005tu}
  N.~Kaiser,
  Phys.\ Rev.\ C {\bf 71} (2005), 068201.

\end{thebibliography}
%


\end{document}